\documentclass[a4paper]{jpconf}
\usepackage{graphicx}

\begin{document}

\title{Measurement-induced entanglement of two superconducting qubits}

\author{Jian Li, K.~Chalapat, G.~S.~Paraoanu}

\address{Low Temperature Laboratory, Helsinki University of Technology,
P.O. Box 5100, FIN-02015 TKK, Finland}

\ead{paraoanu@cc.hut.fi}

\begin{abstract}
We study the problem of two superconducting quantum qubits coupled via a resonator. If only one quanta is present in the system
and the number of photons in the resonator
is measured with a null result, the qubits end up in an entangled Bell state. Here we look at one source of errors in this
quantum nondemolition scheme due to the presence of more than one quanta in the resonator, previous to the measurement.
By analyzing the structure of the conditional Hamiltonian with arbitrary number of quanta,
we show that the scheme is remarkably robust against these type of errors.


\end{abstract}


Superconducting qubits have emerged in the last decade as a reliable benchtop onto which quantum-information experiments
can be implemented. Architectures that demonstrate entanglement at the level of two qubits have been proposed and demonstrated
experimentally (for extensive references, see \cite{us}). Here we concentrate on a scheme that has been proposed in \cite{us}, namely
entanglement of the two qubits {\it via} nondestructive measurements of the number of quanta in third quantum system coupled
to the two qubits.

This third quantum system can be for example a third qubit, a Josephson junction, an electromagnetic resonator, or
a nanomechanical resonator. Schemes such as these have been proposed initially in quantum optics \cite{plenio}, and, together with
interaction-free experiments \cite{interactionfree} and partial-collapse measurements \cite{partialcollapse}, they have no classical analogue
as they explore the consequences of the projective measurements in quantum mechanics. If the number of excitations present in the system is 1, there is no principial difference between
a two-level system and a resonator as the intermediating system. However, in the case of a resonator as a connecting system, it can well be that the
number of excitations present in the system, $n$, is larger than 1. We analyze here precisely this situation.

The Hamiltonian of two qubits (biased at the optimal points) coupled via a resonator is
\begin{eqnarray}
H = -\sum_{j=1,2}\frac{E_{Jj}}{2}\sigma_j^z + \omega_r a^\dag a + i\sum_{j=1,2}g_j(a^\dag \sigma_j^- - a \sigma_j^+) . \nonumber
\end{eqnarray}
In the following, we take the two qubits identical and on-resonance with the coplanar waveguide resonator,
$E_{J1} = E_{J2} = \omega_r$, and the same values for the qubit-resonator coupling constants $g_{1}=g_{2}=g$.

We now consider the situation in which the resonator is out-coupled to a detector. As a detector, one can use a single large Josephson junction
and perform switching-current experiments \cite{paraoanu}. Under the condition that no event is detected by the junction, and
including the
dissipative terms $\Gamma_1 = \Gamma_2 = \Gamma $
for the qubits and the resonator, the conditional non-Hermitian Hamiltonian \cite{carmichael}, in the interaction picture, reads:
\begin{equation}
H_{c} = \omega_{r}\left( a^{\dag}a-\frac{\sigma_{1}^{z}}{2}-\frac{\sigma_{2}^{z}}{2} \right) +ig\left(a^{\dag}\sigma_{1}^{-} +
a^{\dag}\sigma_{2}^{-} - a\sigma_{1}^{+} - a\sigma_{2}^{+}\right)-ika^{\dag}a -i\Gamma\left(\sigma_{1}^{\dag}\sigma_{1}+
\sigma_{2}^{\dag}\sigma_{2}\right)
\end{equation}
\begin{table}
\begin{tabular}{c|c}
\hline
{\rm eigenvalue} & {\rm eigenvector} \\ \hline\hline \\
0 & $\sqrt{\frac{n}{2n-1}}\left(\sqrt{\frac{n-1}{n}}, 0, 0, 1\right)^{T}$ \\  \hline \\
0 & $(0, -1, 1, 0)^{T}$ \\ \hline \\
$-ig\sqrt{2(2n-1)}$ & $\sqrt{\frac{n-1}{2(2n-1)}}\left(-\sqrt{\frac{n}{n-1}}, -i\sqrt{\frac{2n-1}{2(n-1)}}, -i\sqrt{\frac{2n-1}{2(n-1)}}, 1\right)^{T}$  \\  \hline \\
$ig\sqrt{2(2n-1)}$ & $\sqrt{\frac{n-1}{2(2n-1)}}\left(-\sqrt{\frac{n}{n-1}}, i\sqrt{\frac{2n-1}{2(n-1)}}, i\sqrt{\frac{2n-1}{2(n-1)}}, 1\right)^{T}$ \\ \hline
\end{tabular}
\caption{Eigenvectors and eigenvalues of $-iH_{c}^{(n)}$ for $\kappa = \Gamma = 0$.}\label{tabel}
\end{table}
\begin{figure}[htb]
\includegraphics[width=8cm]{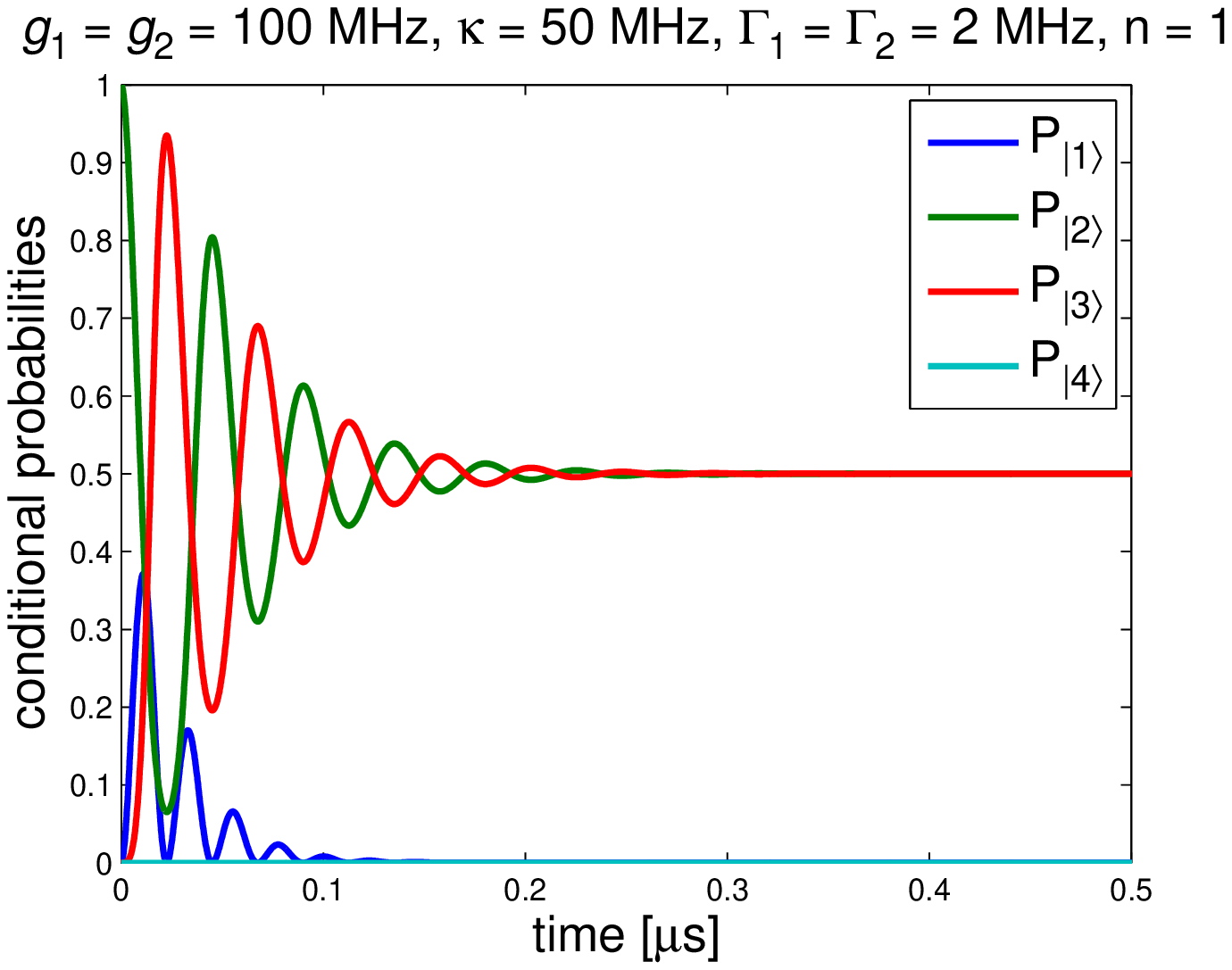}
\includegraphics[width=8cm]{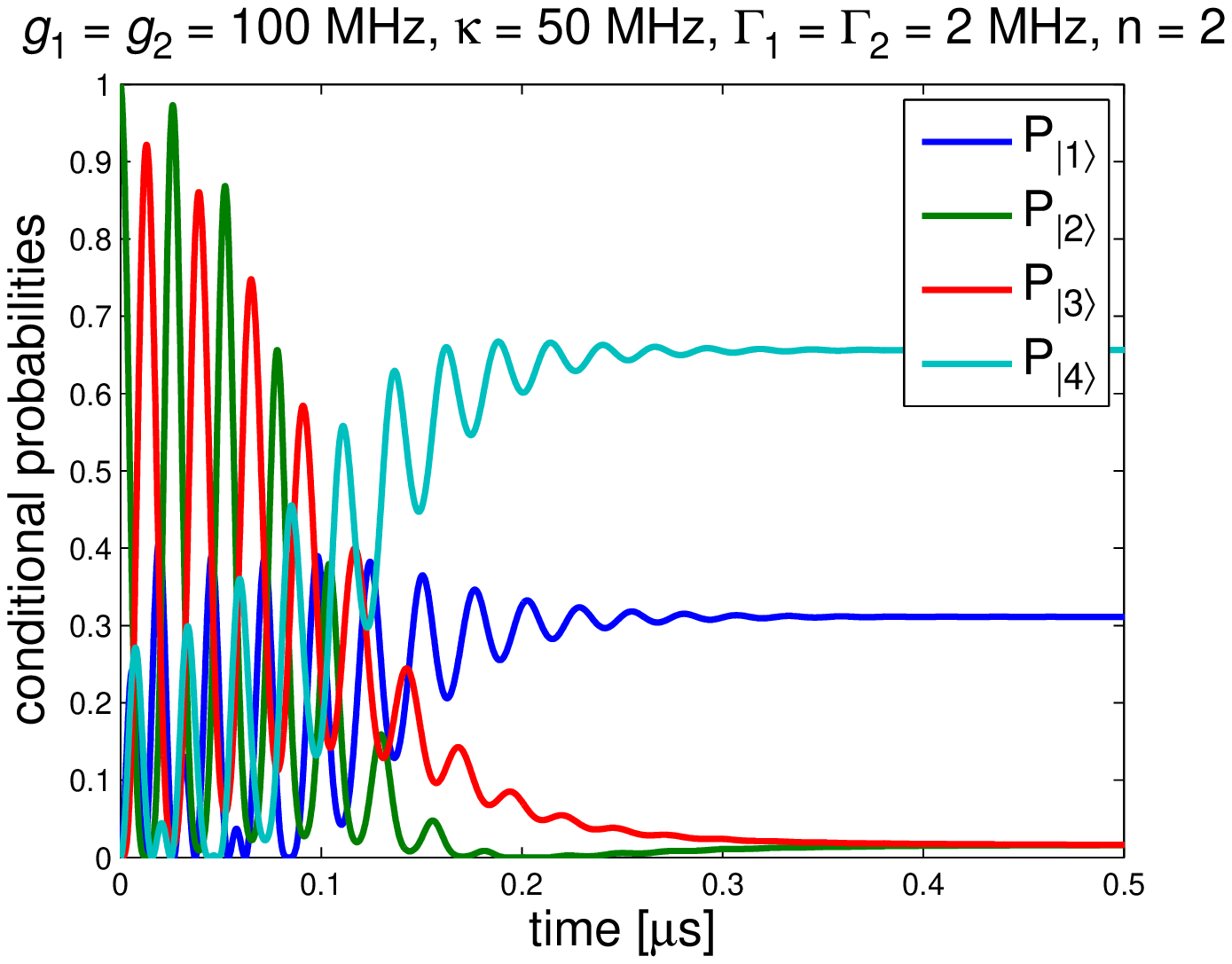}
\includegraphics[width=8cm]{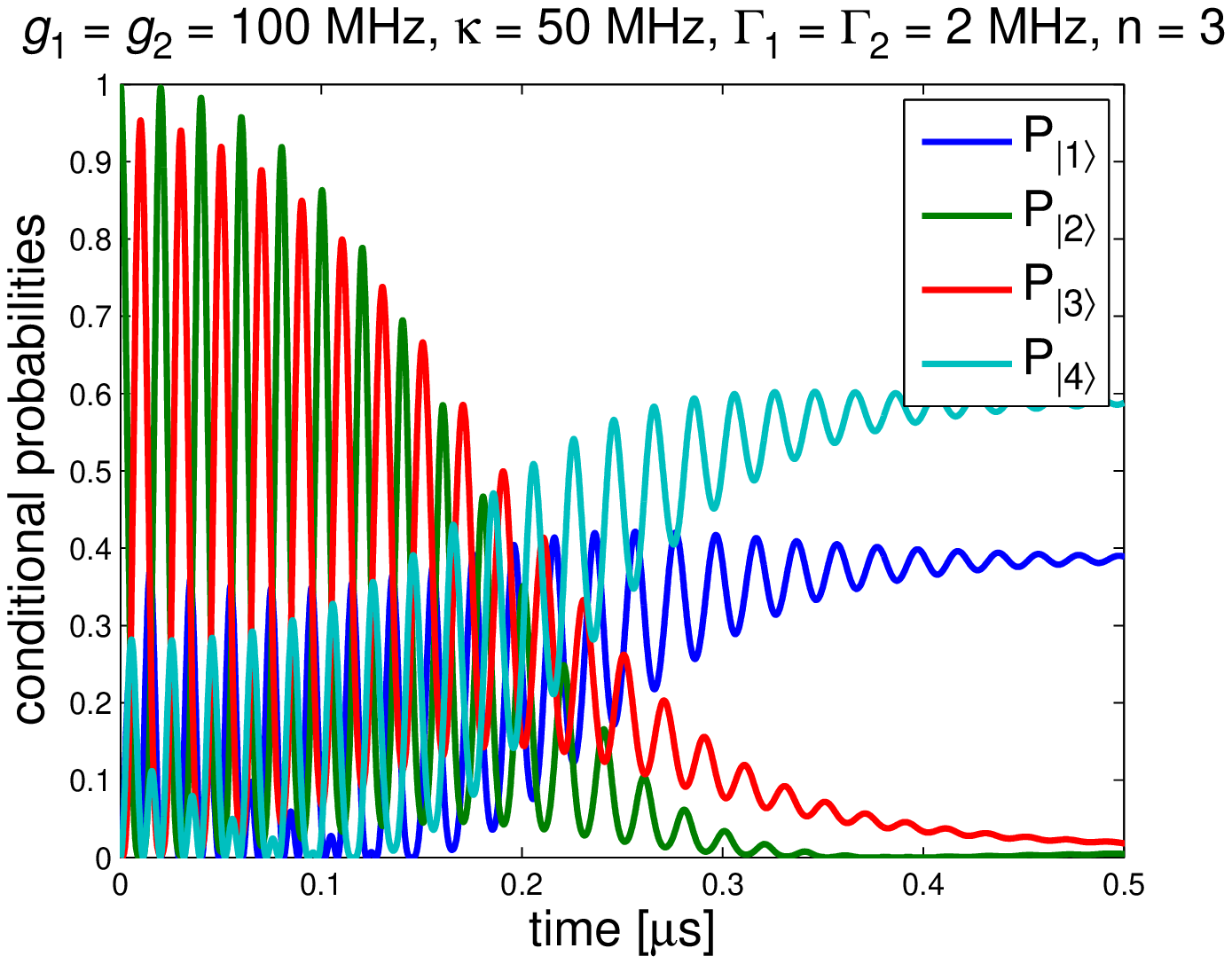}
\includegraphics[width=8cm]{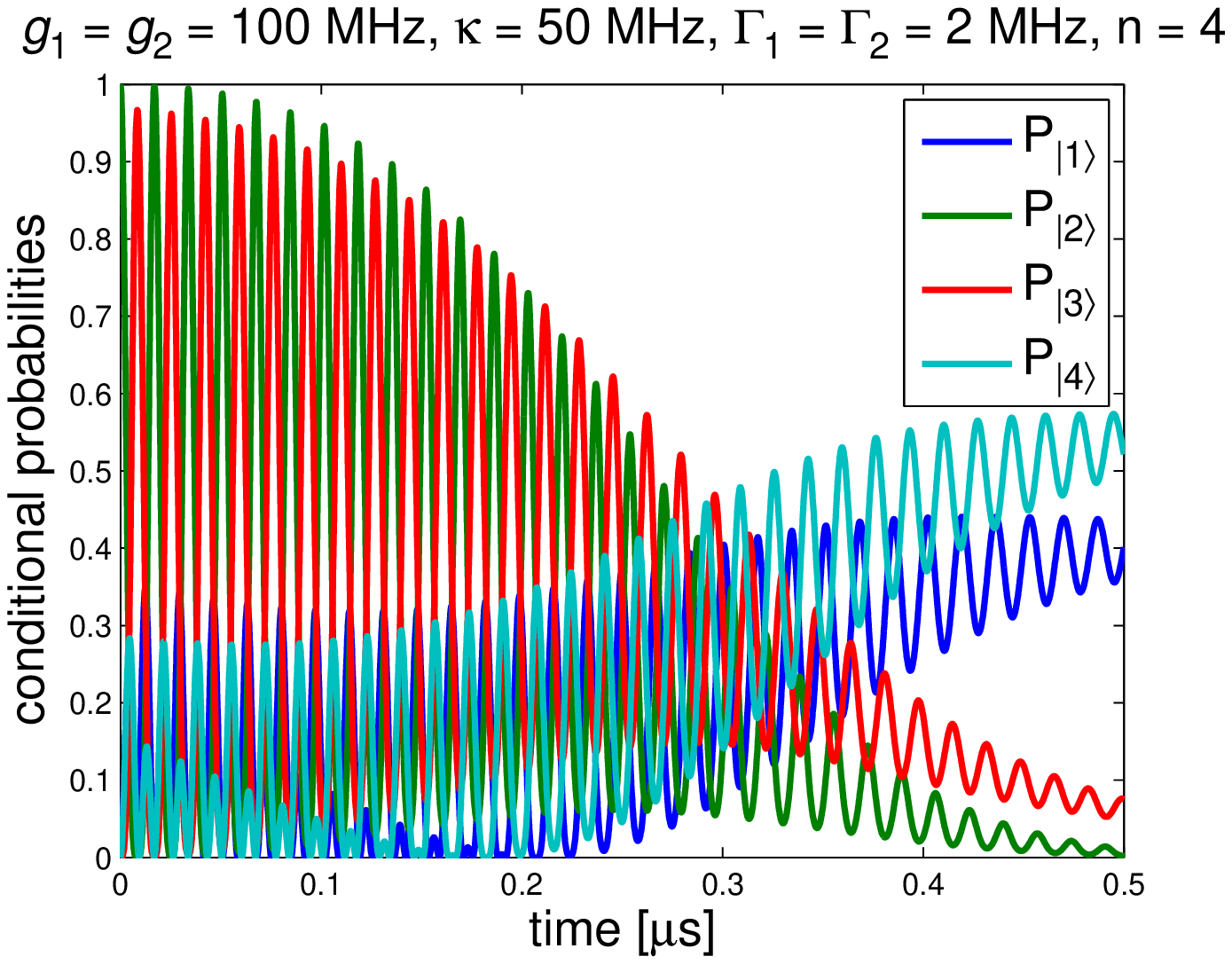}
\caption{The conditional occupation probabilities for the four states  $|\tilde{1},n\rangle$,  $|\tilde{2},n\rangle$,
 $|\tilde{3},n\rangle$, and  $|\tilde{1},n\rangle$, for $n=1,2,3,4$ and initial state $|\tilde{2},n\rangle$.}
\label{fig_cavity_loss}
\end{figure}
We notice that the conditional Hilbert space of the system splits naturally into 4-dimensional subspaces
with $n$ excitations distributed between the two qubits and the resonator.
The Hamiltonian $H_{c}$ projected
onto the subspace spanned by the basis states $|\tilde{1},n\rangle \equiv |n\rangle_\mathrm{p}\otimes|\uparrow\uparrow\rangle$,
$|\tilde{2},n\rangle \equiv |n-1\rangle_\mathrm{p} \otimes |\uparrow\downarrow\rangle$,
$|\tilde{3},n\rangle \equiv |n-1\rangle_\mathrm{p} \otimes |\downarrow\uparrow\rangle$,
and $ |\tilde{4},n\rangle \equiv |n-2\rangle_\mathrm{p} \otimes|\downarrow\downarrow\rangle$
takes the matrix form
\begin{equation}
-iH_{c}^{(n)} = \left[
    \begin{array}{cccc}
        nk & -g\sqrt{n} & -g\sqrt{n}& 0 \\
        g\sqrt{n} & \Gamma + (n-1)k & 0 &-g\sqrt{n-1}\\
        g\sqrt{n}& 0& \Gamma + (n-1)k & -g\sqrt{n-1}\\
        0 &g\sqrt{n-1}&g\sqrt{n-1}&2\Gamma + (n-2)k
        \end{array}
    \right] .
\end{equation}
This non-Hermitian matrix, $-iH_{c}^{(n)}$,  for $k=\Gamma = 0$,  has two eigenvectors with eigenvalue 0, and two
other with nonzero eigenvalue (see Table \ref{tabel}).
We evolve now an initial state with $n$ excitations $|\psi (0)\rangle$ by the conditional Hamiltonian $H_{c}^{(n)}$
and find the state at any time t under the condition that the resonator did not decay, $\exp (-iH_{c}^{(n)}t)|\psi(0)\rangle$.
The results are shown in Fig. (\ref{fig_cavity_loss}).

From Fig. (\ref{fig_cavity_loss}) we see that indeed for $n=1$ the probabilities corresponding
to the states $|\tilde{2}, n=1\rangle$ and $|\tilde{3}, n=1\rangle$ are equal. It is in fact known \cite{us,plenio},
that, if the system starts in the states $|\tilde{3}, n=1\rangle$ or $|\tilde{2}, n=1\rangle$
in the
asymptotic state the qubits will be projected to a maximally entangled Bell state,
$|\tilde{3}, n=1\rangle-|\tilde{2}, n=1\rangle +
=(1/\sqrt{2}) |0\rangle_{p} \otimes (|\uparrow\downarrow\rangle - |\downarrow\uparrow\rangle )$
(we have verified the also the minus sign numerically).
This corresponds to the second eigenvalue of the nondissipative Hamiltonian from Table \ref{tabel}.

In the case of $n\neq 1$ photons a completely different state results, namely the asymptotic state corresponds
to the first eigenvalue in Table \ref{tabel}. This statement is valid no matter which of the four states (or combinations of them)
$|\tilde{1}, n\rangle$, $|\tilde{2}, n\rangle$, $|\tilde{3}, n\rangle$, $|\tilde{4}, n\rangle$
is taken as the initial one, although of course the intermediate-time evolution is different.

From Fig. (\ref{fig_cavity_loss}) we see that the ratios of the two
surviving probabilities are indeed $P_{|\tilde{1},n\rangle}/P_{|\tilde{4},n\rangle}=(n-1)/n$; we have also checked numerically the signs.
The asymptotic state is then
\begin{equation}
|\psi_{asym} \rangle = \sqrt{\frac{n}{2n-1}}\left( \sqrt{\frac{n-1}{n}}|\tilde{1}, n\rangle + |\tilde{4}, n\rangle \right)
\end{equation}
Interestingly, in the limit of large $n$ in which we can take $|n\rangle_{p}\approx|n-2\rangle_{p}$ and factor out the resonator, this
state results in another two-qubit Bell state, $(|\uparrow\uparrow\rangle + |\downarrow\downarrow\rangle )/\sqrt{2}$.
\begin{figure}[htb]
\includegraphics[width=8cm]{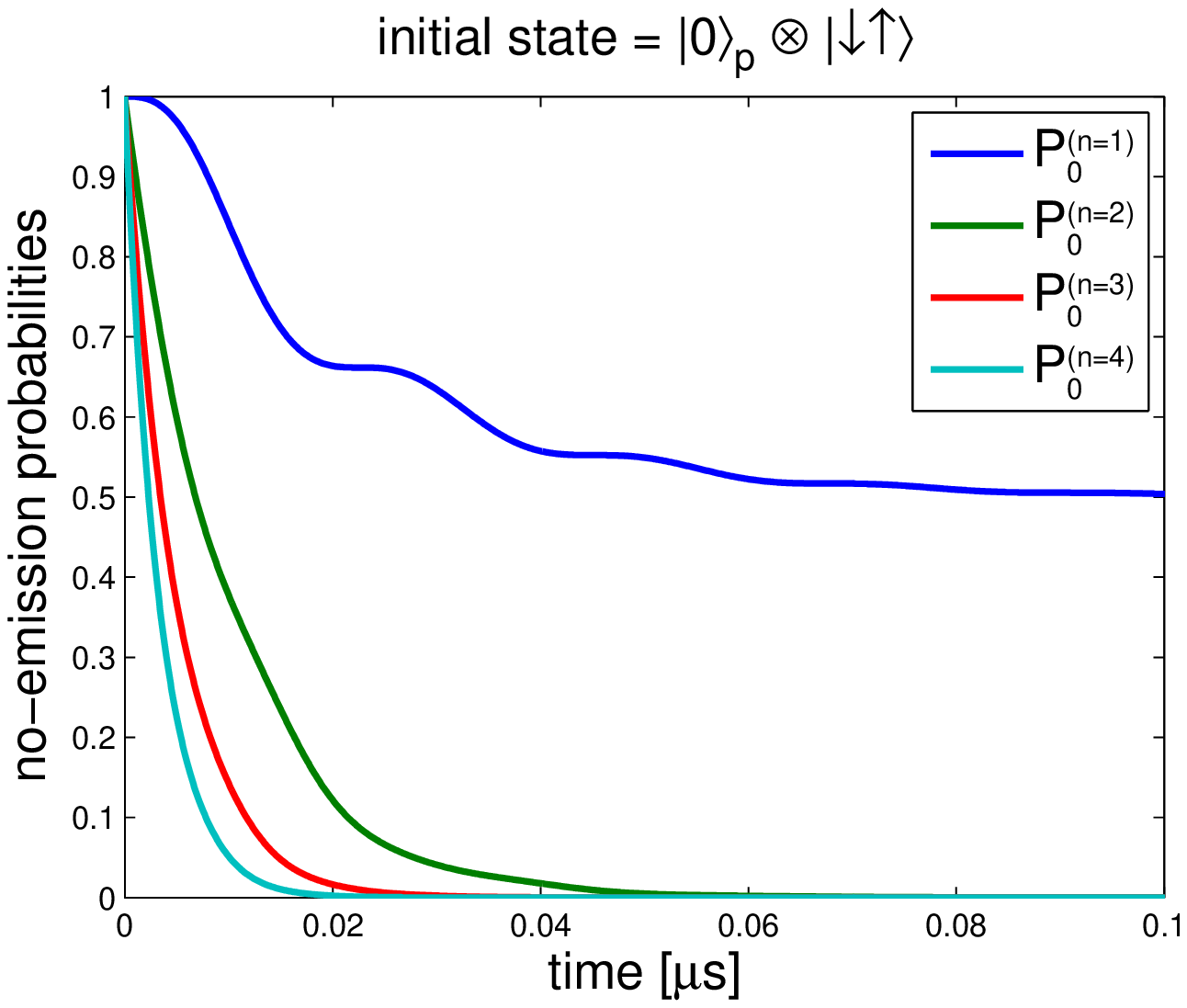}
\includegraphics[width=8cm]{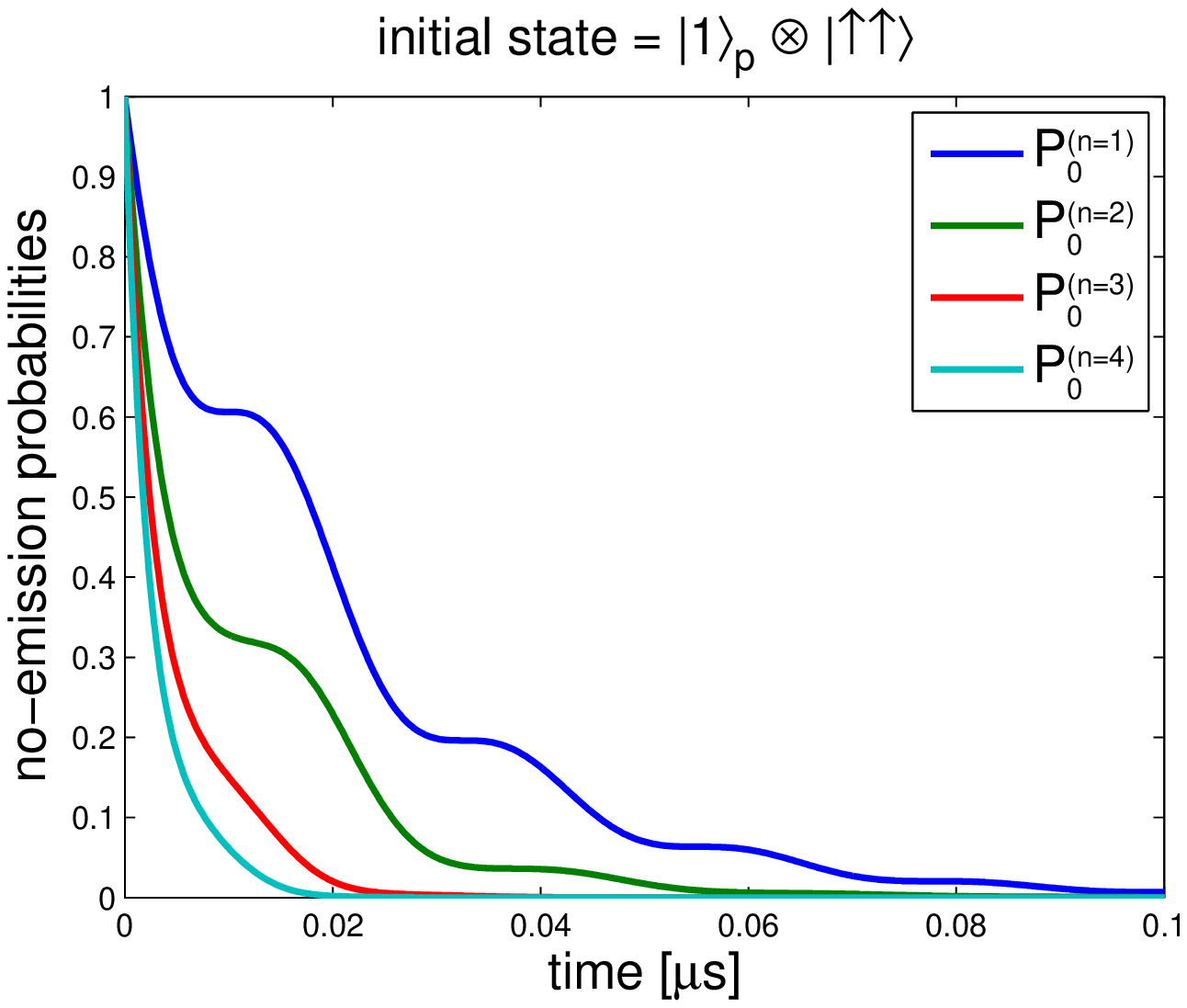}
\caption{The probability of no photon emission in the limit $\Gamma^{-1}\gg k^{-1}$ and for different number of excitations.
The graph on the left corresponds to an initial state $|\tilde{2},n=1\rangle$,
while that on the right to $|\tilde{1},n=1\rangle
$. The other parameters were $g_1 = g_2 = 100$ MHz, $\kappa = 50$ MHz, $\Gamma_1 = \Gamma_2 = 0$ .}
\label{norms}
\end{figure}
It is intuitively clear that due to the presence of extra excitations in the system the probability of decay will increase
and there will be no asymptotic stationary state as in the case $n=1$. To understand this quantitatively, we
calculate the probability to have no photon emission until time t. This is given by the norm of the wavefunction evolved with
the non-Hermitiam conditional Hamiltonian with $n$ excitations,
\begin{equation}
P_{0}^{(n)}(t)=||e^{-i H_{c}^{(n)}t}|\psi(0)\rangle||^2.
\end{equation}
This probability is plotted in Fig. \ref{norms} for $n=1,2,3,4$. To make the interpretation straigthforward, we took $\Gamma_{1}=
\Gamma_{2} = \Gamma = 0$, in other words the only possibility of decay is through the resonator. We see that for $n=1$ and
an initial state $|\tilde{2},n=1\rangle$ (or equivalently $|\tilde{3},n=1\rangle$)
the probability
$P_{0}^{(n=1)}$ goes asymptotically to
0.5, as it should in the case $k\gg\Gamma$, as was found in \cite{us,plenio}, corresponding to an eigenvalue of the form
$(|\uparrow\downarrow\rangle + |\downarrow\uparrow\rangle)/\sqrt{2}$. For the other possible initial state with $n=1$,
namely $|\tilde{1},n=1\rangle$, the probability decays fast, therefore the errors introduced by these states will be asymptotically
suppressed.

For $n\neq 1$ the probability of no decay goes very fast to zero: this means that the efficiency of the scheme is much smaller
than in the case $n=1$, in the sense that there will be much less "favorable" events in which no decay of the resonator is detected.
To make things more formal, suppose we start with an initial mixture of states $|\tilde{2},n=1\rangle$
(desirable, probability $P^{(0)}$) and thermal excitations or other kind of errors and fluctuations giving $n \neq 1$ excitations in the system
with probabilities $P_{ex}^{(n)}$. The explicit form of this initial state does not matter, because the asymptotic result
is the same:  the asymptotic state
will be a mixture
of the Bell state we want to obtain,
$(|\uparrow\downarrow\rangle + |\downarrow\uparrow\rangle)/\sqrt{2}$, which comes with probability $1/2$
and states corresponding to  $|\psi_{asym}\rangle$, which come with (the much smaller) probabilities
$P_{0}^{(n)}$,
\begin{eqnarray}
\rho_{asym} &=& \frac{0.25 P^{(0)}}{0.5 P^{(0)} + \sum_{m}P_{0}^{(m)}P_{ex}^{(m)} }
(|\uparrow\downarrow\rangle + |\downarrow\uparrow\rangle )(\langle\uparrow\downarrow | +
\langle\downarrow\uparrow |) \otimes |0\rangle_{pp}\langle 0 |\nonumber \\
& &+
\sum_{n} \frac{P_{0}^{(n)}P_{ex}^{(n)}}{0.5 P^{(0)} + \sum_{m}P_{0}^{(m)}P_{ex}^{(m)} }
|\psi_{asym}\rangle\langle \psi_{asym}|) .\nonumber
\end{eqnarray}
In conclusion, irrespective to the initial weights $P^{(0)}$, $P_{ex}^{(n)}$,
 this density matrix approaches exponentially fast the desirable Bell state, due to the fast decay of the probabilities
$P_{0}^{(n)}$. Of course, the initial state and $P^{(0)}$, $P_{ex}^{(n)}$ do have a role to play in the efficiency of this
process (how many favorable non-counting events we get), but not in the structure of the final state under the condition
of no de-excitation events detected.

\label{conclude}
\ack

This work was supported by the Academy of Finland (Projects No. 00857, No. 7111994, and No. 7118122).

\end{document}